%% file: skeleton.tex
\documentclass[a4paper,11pt]{article}
\usepackage{pos}
\usepackage{amsmath}

\usepackage{amssymb}
\usepackage{anyfontsize}
\usepackage{subfigure}
\DeclareMathAlphabet{\mathcal}{OMS}{cmsy}{m}{n}

\graphicspath{{./}}

\input{newcommands.tex}

\title{Improved analysis of nucleon isovector charges and twist-2 matrix elements on CLS $N_f=2+1$ ensembles}
\ShortTitle{Improved analysis of nucleon isovector charges and twist-2 matrix elements}

\author*[a]{Konstantin~Ottnad}
\author[b,c]{Dalibor~Djukanovic}
\author[d]{Tim~Harris}
\author[a,b,c]{Harvey~B.~Meyer}
\author[a]{Georg~von~Hippel}
\author[a,b,c]{Hartmut~Wittig}

\affiliation[a]{PRISMA\textsuperscript{+} Cluster of Excellence and Institut f\"ur Kernphysik, Johannes Gutenberg-Universität Mainz, \\
Johann-Joachim-Becher-Weg 45, 55099 Mainz, Germany}

\affiliation[b]{Helmholtz Institute Mainz, Johannes Gutenberg-Universität Mainz, 55099 Mainz, Germany}

\affiliation[c]{GSI Helmholtzzentrum für Schwerionenforschung, Planckstraße 1, 64291 Darmstadt, Germany}

\affiliation[d]{School of Physics and Astronomy, University of Edinburgh, \\
Peter Guthrie Tait Road, Edinburgh, EH9 3JZ, UK}

\emailAdd{kottnad@uni-mainz.de}

\abstract{Preliminary results are presented for nucleon isovector charges and twist-2 matrix elements which have been obtained employing an improved analysis strategy to deal with excited-state contamination. The set of CLS $N_f=2+1$ gauge ensembles in this study has been extended compared to our 2018 calculation \cite{Harris:2019bih}, including an ensemble at physical quark masses. Besides the addition of new ensembles, the number of gauge configurations and measurements has been increased on several of the existing ensembles and the analysis has been extended to include additional source-sink separations. The ensembles cover a range of the light quark mass corresponding to $M_\pi\approx 0.130\,\mathrm{MeV} \ldots 350\,\mathrm{MeV}$, four values of the lattice spacing $a\approx0.05\,\mathrm{fm}\ldots0.09\,\mathrm{fm}$ and a large range of volumes. Results at the physical point are computed for each observable from a combined chiral, continuum and finite-volume extrapolation.}

\FullConference{
 The 38th International Symposium on Lattice Field Theory, LATTICE2021
  26th-30th July, 2021
  Zoom/Gather@Massachusetts Institute of Technology
}

\begin{document}
\maketitle

\section{Introduction}
  In this contribution we report on an ongoing effort of the Mainz group to update and improve our results for isovector nucleon charges and twist-2 matrix elements that have been published a couple of years ago in Ref.~\cite{Harris:2019bih} (see also Refs.~\cite{Ottnad:2017mzd,Ottnad:2018fri}). The changes and improvements include, but are not limited to: additional gauge ensembles, lighter quark masses, larger statistics and a different procedure to treat contamination from excited states. The latter becomes necessary as the excited-state spectrum becomes more dense towards physical light quark mass, hence requiring better control over the resulting contamination. \par

More explicitly, we aim at computing forward nucleon matrix elements at zero momentum transfer with initial and final state produced at rest
\begin{equation}
 \bra{N(s_f)} \mathcal{O}^X_{\mu_1...\mu_n} \ket{N(s_i)} = \bar{u}(s_f) W^X_{\mu_1...\mu_n}  u(s_i) \,,
 \label{eq:matrix_element}
\end{equation}
where $u(s_i)$, $\bar{u}(s_f)$ are Dirac spinors with initial (final) state spin $s_i$ ($s_f$). For the operator insertion we consider the quark isovector combination of local, dimension-three operators
\begin{equation}
 \mathcal{O}^A_\mu=\bar{q} \gamma_\mu \gamma_5 q, \quad \mathcal{O}^S=\bar{q}q , \quad \mathcal{O}^T_{\mu\nu} = \bar{q} i \sigma_{\mu\nu} q \,,
 \label{eq:local_operators}
\end{equation}
and the one-derivative, dimension-four operators
\begin{equation}
  \mathcal{O}^{vD}_{\mu\nu}=\bar{q} \g{\l\{\mu\r.} \DBF{\l.\nu\r\}} q \,, \qquad \mathcal{O}^{aD}_{\mu\nu}=\bar{q} \g{\l\{\mu\r.} \g{5} \DBF{\l.\nu\r\}} q \,, \qquad \mathcal{O}^{tD}_{\mu\nu\rho}=\bar{q} \sigma_{\l[\mu\l\{\nu\r.\r]} \DBF{\l.\rho\r\}} q \,,
 \label{eq:twist2_operators}
\end{equation}
where $\DBF{\mu}=\frac{1}{2} (\DF{\mu}-\DB{\mu})$ and the shorthands $\{...\}$ and $[...]$ denote symmetrization over indices with subtraction of the trace and anti-symmetrization, respectively. The operator-dependent form factor decompositions $W^X_{\mu_1...\mu_n}$ take a simple form at zero momentum transfer and yield direct access to the charges $g_{A,S,T}^{u-d}$ and the average quark momentum fraction $\avgx{-}{}$ as well as the helicity and transversity moments $\avgx{-}{\Delta}$ and $\avgx{-}{\delta}$. \par

The lattice calculation of these matrix elements is carried out by computing appropriately spin-projected two- and three-point functions $C^\mathrm{2pt}(\tsep)$ and $C^X_{\mu_1...\mu_n}(\tins, \tsep)$, where $X=A,S,T,...$ labels the operator insertion. Initial and final state are produced at rest in both cases. The relevant Euclidean time separations are the source-sink separation $\tsep=t_f-t_i$ and the insertion time $\tins=t - t_i$, where $t_i$ ($t_f$) denotes the source (sink) time and $t$ the time position of the insertion. The desired matrix elements are obtained at asymptotically large Euclidean time separations from the ratio
\begin{equation}
 R^X(\tins, \tsep) = \frac{C^X_{\mu_1...\mu_n}(\tins, \tsep)}{C^\mathrm{2pt}(\tsep)} \,,
 \label{eq:ratio}
\end{equation}
that is required to cancel unknown overlap factors. However, the time-separations $\tsep\gg 1\fm$ that are needed to ensure ground state dominance are impossible to achieve even with state-of-the-art lattice methods, hence the use of additional, dedicated  techniques to reduce excited-state contamination remains mandatory. Further details on the lattice setup and in particular changes compared to the setup used in Ref.~\cite{Harris:2019bih} concerning ensembles, statistics, available source-sink separations etc. are detailed in section~\ref{sec:setup}. Regarding our analysis strategy for taming excited states we have implemented major changes over the previous study, and the subject is discussed in some detail in section~\ref{sec:excited_states}. The procedure for physical extrapolations, on the other hand, remains the same as before and is only briefly explained in section~\ref{sec:physical_extrapolations} where preliminary results are given and some future plans and directions are outlined. \par

\section{Lattice setup}
\label{sec:setup}
Our lattice simulations are carried out on gauge ensembles with $N_f=2+1$ flavors of non-perturbatively $\mathcal{O}(a)$-improved Wilson quarks that have been generated by the Coordinated Lattice Simulations (CLS) initiative \cite{Bruno:2014jqa}. The simulations employ the tree-level Symanzik gauge action and a twisted mass regulator \cite{Luscher:2012av} to suppress exceptional configurations. The ensembles used in this study are listed in Tab.~\ref{tab:ensembles}. While most of them have been generated with open boundary conditions (oBC) in time to reduce the effect of topological charge freezing~\cite{Luscher:2011kk} at finer lattice spacing, some new ensembles with periodic boundary conditions (pBC) have been included as well. \par

\begin{table}[!t]
 \centering
 \begin{tabular}{ccrrccrrcc}
  \hline\hline
  ID$^\mathrm{BC}$ & $\beta$ & T/a & L/a & $M_\pi / \gev$ & $M_\pi L$ & $N_\mathrm{conf}$ & $N_\mathrm{meas}$ & $\tseplo/\fm$ & $\tsephi/\fm$ \\
  \hline\hline
  C101$^o$ & 3.40 &  96 & 48 & 0.2250(12) & 4.73 & 2000 &  64000 & 0.35 & 1.38 \\
  H105$^o$ & 3.40 &  96 & 32 & 0.2805(27) & 3.93 & 1027 &  49296 & 0.35 & 1.38 \\
  H102$^o$ & 3.40 &  96 & 32 & 0.3543(11) & 4.96 & 2005 &  32080 & 0.35 & 1.38 \\
  \hline
  D450$^p$ & 3.46 & 128 & 64 & 0.2163(07) & 5.35 &  500 &  64000 & 0.31 & 1.53 \\
  N451$^p$ & 3.46 & 128 & 48 & 0.2860(05) & 5.31 & 1011 & 129408 & 0.31 & 1.53 \\
  S400$^o$ & 3.46 & 128 & 32 & 0.3496(11) & 4.33 & 2873 &  45968 & 0.31 & 1.53 \\
  \hline
  E250$^p$ & 3.55 & 192 & 96 & 0.1299(09) & 4.06 &  250 &  64000 & 0.26 & 1.41 \\
  D200$^o$ & 3.55 & 128 & 64 & 0.2024(08) & 4.22 & 2000 &  64000 & 0.26 & 1.41 \\
  N200$^o$ & 3.55 & 128 & 48 & 0.2811(09) & 4.39 & 1712 &  20544 & 0.26 & 1.41 \\
  S201$^o$ & 3.55 & 128 & 32 & 0.2924(16) & 3.05 & 2093 &  66976 & 0.26 & 1.41 \\
  N203$^o$ & 3.55 & 128 & 48 & 0.3459(08) & 5.41 & 1543 &  24688 & 0.26 & 1.41 \\
  \hline
  J303$^o$ & 3.70 & 192 & 64 & 0.2596(08) & 4.19 & 1073 &  17168 & 0.20 & 1.40 \\
  N302$^o$ & 3.70 & 128 & 48 & 0.3485(08) & 4.22 & 2201 &  35216 & 0.20 & 1.40 \\
  \hline\hline
 \end{tabular}
 \caption{List of CLS gauge ensembles used in this work with the respective choice of boundary conditions in time (``$o$'': open, ``$p$'': periodic) and the values of $\beta$, $T/a$ and $L/a$. The measured pion masses are given in physical units. In addition, the values of $M_\pi L$ have been included, as well as the number of gauge configurations $N_\mathrm{conf}$ and the number of measurements $N_\mathrm{meas}$ which refers either to $\tsep\geq 1\fm$ for ensembles with open boundaries or to the largest value of $\tsep$ otherwise. The last two columns show the lower and upper bound of the available source-sink separations $\tseplo$ and $\tsephi$ in physical units.}
 \label{tab:ensembles}
\end{table}

\subsection{Comparison with previous analysis}
Compared to the analysis in Ref.~\cite{Harris:2019bih} numerous changes have been implemented. First of all, three ensembles with pBC have been added, i.e. D450, N451 and, most notably, the E250 ensemble at physical quark masses. Note that N451 replaces N401 with a change of boundary conditions (i.e. pBC vs oBC) while sharing the same simulation input parameters. Besides, correlation functions on H102 have been recomputed to include twist-2 operator insertions making them consistently available for all ensembles. As a byproduct we have increased effective statistics on H102 by a factor of four at source-sink separations $\tsep\geq 1\fm$. Moreover, the number of gauge configurations has been increased on ensembles D200, J303, N302 and S400 by up to a factor of two. \par

Another important change concerns the generation of three-point function data at values of $\tsep<1\fm$. These are required for our excited-state analysis that will be explained in Section~\ref{sec:excited_states}. In general, the smallest, available source-sink separation now corresponds to $\tsep/a=4$, and data for all even values of $\tsep/a$ between $\tseplo$ and $\tsephi$ have been computed, cf. Table~\ref{tab:ensembles}. Triggered by the inclusion of smaller values of $\tsep$, we have modified the source setup used for the computation of two- and three-point functions due to the expected exponential improvement of the signal-to-noise ratio. This will be detailed in the next subsection, together with the strategy for ensembles with pBC. Concerning additional source-sink separations, we have also added three-point function data at $\tsep=28a$ for J303, such that data up to $\tsep\geq 1.4\fm$ are consistently available for all ensembles. \par

Finally, we remark that our procedure for renormalization remains unchanged. In particular, we use $Z_A$ from the Schr\"odinger functional approach for the axial vector matrix elements at all four values of $\beta$; cf. Refs.~\cite{DallaBrida:2018tpn,Korcyl:2016ugy,Gerardin:2018kpy}. For the remaining matrix elements we use the renormalization constants that have been determined in Ref.~\cite{Harris:2019bih} to which we also refer for further details and references. \par

\subsection{Source setup and scaling of effective statistics}
The computation of two- and three-point functions is carried out using the truncated solver method \cite{Bali:2009hu,Blum:2012uh,Shintani:2014vja} to reduce the computational cost by a factor of $\sim2$ to 3. This now includes H102 for which only exact solves had previously been available. For the oBC ensembles that have already been used in Ref.~\cite{Harris:2019bih} the three-point function data at $\tsep \geq 1\fm$ has been generated on a fixed set of source positions on a single timeslice in the bulk of the lattice, regardless of the actual value of $\tsep$. The resulting number of measurements is given in Table~\ref{tab:ensembles}. However, keeping the same setup for data at (much) smaller values of $\tsep$ would give them too much weight in fits. In order to remedy this issue, the number of sources has been reduced by a factor of two for each or every second step in decreasing $\tsep$. The spatial source positions for these additional values of $\tsep$ have been randomly drawn without replacement on a fixed timeslice for each configuration on oBC ensembles, subject to the usual constraints caused by the combination of the truncated solver method and the SAP preconditioning \cite{Luscher:2003vf,vonHippel:2016wid}. In case of H102 the sources have been randomly sampled in the same way also for $\tsep \geq 1\fm$, and we intend to keep this kind of setup for future runs on oBC ensembles instead of using fixed source positions. An important side effect of scaling down the number of sources at decreasing values of $\tsep$ is the reduction in computational cost compared to the setup originally used for the production of data at $\tsep\geq 1\fm$ with full statistics. \par

For pBC ensembles, on the other hand, the sources are randomly sampled from the entire lattice volume and otherwise adhere to the previously mentioned constraints. In this case $N_\mathrm{meas}$ refers to $\tsephi$ only, as we have scaled statistics down across the full range of source-sink separations, resulting in an even smoother signal-to-noise behavior. In general, sampling the entire lattice volume allows us one to place many more sources before saturation occurs, which explains the larger $N_\mathrm{meas}/N_\mathrm{conf}$ on these ensembles. For example, on N451 the value of $N_\mathrm{meas}$ has been increased by an order of magnitude compared to N401 in Ref.~\cite{Harris:2019bih}, whereas $N_\mathrm{conf}$ has been increased by less than a factor of two.\par

\section{Excited state treatment}
\label{sec:excited_states}
Excited states are typically the dominant systematic effect in lattice calculations of nucleon matrix elements. This is caused by the exponential signal-to-noise problem that renders the use of sufficiently large Euclidean time-separations impossible for computations of the ratio in Eq.~(\ref{eq:ratio}) at reasonable computational cost. Therefore, various methods have been developed to achieve additional suppression of contamination by excited states. For a recent review on the subject we refer to Ref.~\cite{Ottnad:2020qbw}. In this study we focus on implementations of the summation method \cite{Maiani:1987by,Dong:1997xr,Capitani:2012gj}, which we find to have several advantages over the simultaneous two-state fits that we have mainly used in Ref.~\cite{Harris:2019bih} when applied to our now extended and improved set of lattice data. \par

In order to obtain our summation method fit models, we take the expressions for the two-state truncation of nucleon two- and three-point functions at vanishing momentum transfer, i.e.
\begin{align}
 C^\mathrm{2pt}(\tsep) =& \l|A_0\r|^2 e^{-m_0 \tsep} + \l|A_1\r|^2 e^{-m_1 \tsep} + ... \,, \label{eq:2pt_two_state_truncation} \\[0.25em]
 C^X_{\mu_1...\mu_n}(\tins, \tsep) \notag =& |A_0|^2 M_{00} e^{-m_0\tsep} + A_0 A_1^* M_{01} e^{-m_0(\tsep-\tins)} e^{-m_1\tins} \notag \\
  &+ A_1 A_0^* M_{10} e^{-m_1(\tsep-\tins)} e^{-m_0\tins} + |A_1|^2 M_{11} e^{-m_1\tsep} + ... \,.
 \label{eq:3pt_two_state_truncation}
\end{align}
and plug them into the ratio in Eq.~(\ref{eq:ratio}), which yields
\begin{equation}
 R^X(\tins, \tsep) = \frac{M_{00} + M_{01} \frac{A_1^*}{A_0^*} e^{-\Delta\tins} + M_{01} \frac{A_1}{A_0} e^{-\Delta(\tsep-\tins)} + M_{11} \frac{|A_1|^2}{|A_0|^2} e^{-\Delta\tsep}}{1 + \frac{|A_1|^2}{|A_0|^2} e^{-\Delta\tsep}} \,,
 \label{eq:ratio_two_state_truncation_pi0pf0}
\end{equation}
where the leading gap $\Delta=m_1-m_0$ has been introduced and we made use of the fact that $M_{01}=M_{10}$ at zero-momentum transfer. The expression for the two-state truncation of the summed ratio $S(\tsep, \tex)=\sum_{\tins=\tex}^{\tsep-\tex} R(\tins, \tsep)$ is obtained from an expansion in $e^{-\Delta\tsep}$ around zero and keeping only terms up to and including $\mathcal{O}(e^{-\Delta\tsep})$
\begin{align}
 S(\tsep,\tex) = & M_{00}(\tsep - 2\tex + a) + 2 \tilde{M}_{01} \frac{ e^{-\Delta\tex} - \bigl( e^{\Delta(\tex-a)} + \frac{|A_1|^2}{|A_0|^2} e^{-\Delta\tex}\bigr) e^{-\Delta\tsep}}{1-e^{-\Delta a}} \notag \\
                 & +\tilde{M}_{11} e^{-\Delta\tsep} (\tsep - 2\tex + a) + \mathcal{O}(e^{-2\Delta\tsep}) \,.
\end{align}
Here we have defined $\tilde{M}_{01}=2\mathrm{Re}\l[ A_1/A_0\r] M_{01}$ and $\tilde{M}_{11}=|A_1|^2/|A_0|^2 (M_{11}-M_{00})$.
However, at our current level of statistics we find that terms $\sim \frac{|A_1|^2}{|A_0|^2}$ are not constrained by the data. Therefore, we decided to drop them, leading to our preferred fit model
\begin{equation}
 S(\tsep, \tex=a) = M_{00}(\tsep - a) + 2 \tilde{M}_{01} \frac{ e^{-\Delta a} - e^{-\Delta\tsep} }{1-e^{-\Delta a}} \,.
 \label{eq:summation_two_state}
\end{equation}
where we have set $\tex=a$. The expression is fitted simultaneously in all six observables with $\Delta$ as a common fit parameter. For the lower bound of the fit range we demand $M_\pi\tsepmin\geq0.5$, with $\tsepmin$ being further increased for ensembles with high statistics to achieve reasonable fit quality. A less constraining variation of the two-state fit model is given by \footnote{A fourth term $\sim e^{-\Delta\tsep}(\tsep-2\tex+a)$ is again dropped as it is proportional to $\sim \frac{|A_1|^2}{|A_0|^2}$.}
\begin{equation}
 S(\tsep, \tex=a) = a_0 + a_1 (\tsep - a) + a_2 e^{-\Delta\tsep};
 \label{eq:summation_two_state_nonuniversal}
\end{equation}
where $c_1=M_{00}$ and $c_0$ collects constant contribution from all higher states. This is similar to the standard form of the summation method
\begin{equation}
 S(\tsep, \tex=a) = b_0 + b_1 (\tsep - a) \,,
 \label{eq:summation_single_state}
\end{equation}
where $b_1=M_{00}$. Unlike Eqs.~(\ref{eq:summation_two_state}),~(\ref{eq:summation_two_state_nonuniversal}) the standard form of the summation method exhibits no common parameters and is thus fitted to each observable individually. We include it for comparison purposes and impose a more restrictive cut in $\tsepmin$, i.e. $M_\pi \tsep^{\min} \geq 0.7$. \par

The summation-based approach has several features that make it an attractive alternative to e.g. the simultaneous two-state ratio fit model that has been used in Ref.~\cite{Harris:2019bih}
\begin{equation}
 R(\tins, \tsep) = c_0 + c_1 (e^{-\Delta\tins} - e^{-\Delta(\tsep-\tins)}) + c_2 e^{-\Delta\tsep} \,. \notag
 \end{equation}
where we have $c_0=M_{00}$ and data are fitted simultaneously in $\tsep\geq\tsepmin$ and $\tins\in\left[\tinsmin,\tsep/2\right]$ with $\tinsmin=\tsepmin/2$. The leading correction is of $\mathcal{O}(e^{-\Delta \tsepmin/2})$ in this model, while for the approach based on the summation method it is of $\mathcal{O}(e^{-\Delta \tsepmin})$. This enhanced suppression is an important advantage of the summation method as it allows to include more precise and / or numerically cheaper data at smaller values of $\tsep$, similar to what has been argued in ref.~\cite{Chang:2018uxx}. In particular, demanding comparable suppression by imposing a constraint on $\tinsmin$ yields $M_\pi \tinsmin\geq0.5$ which is equivalent to $M_\pi \tsepmin\geq1$. On e.g. E250 this would imply $\tsepmin\approx1.5\fm$, effectively excluding all data from the fit. Last but not least, the summation method fit models involve fewer degrees of freedom and hence smaller covariance matrices, which greatly improves stability. \par

\begin{figure}[thb]
 \centering
 \includegraphics[totalheight=0.175\textheight]{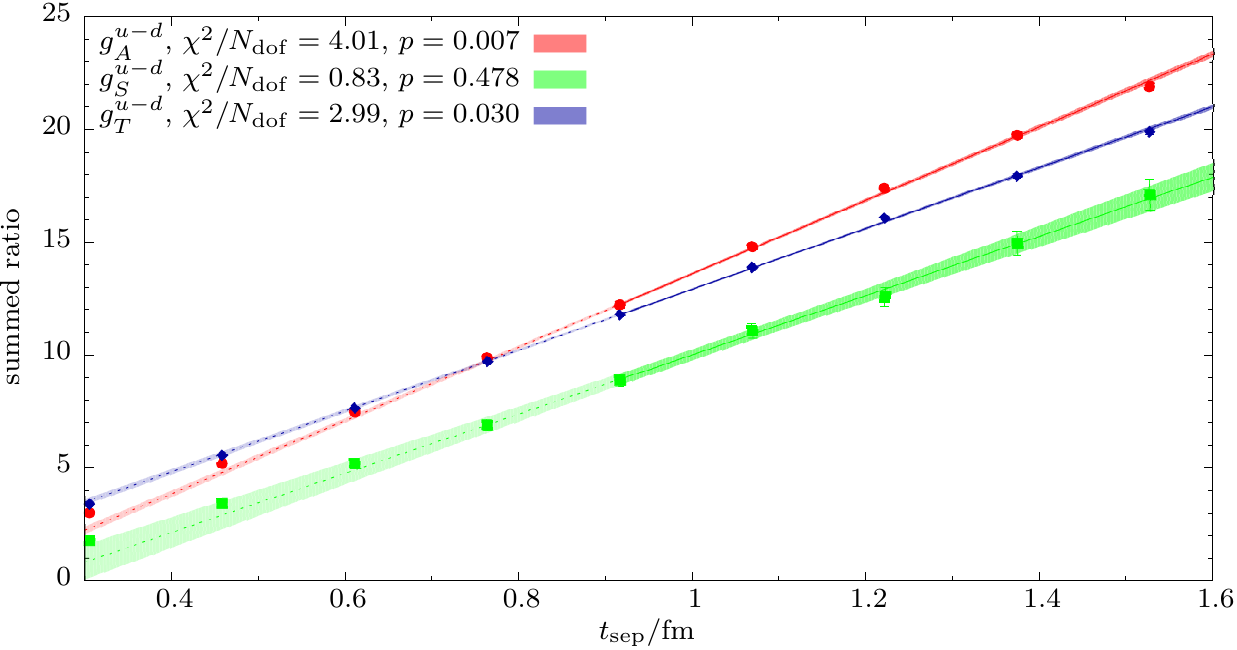}
 \includegraphics[totalheight=0.175\textheight]{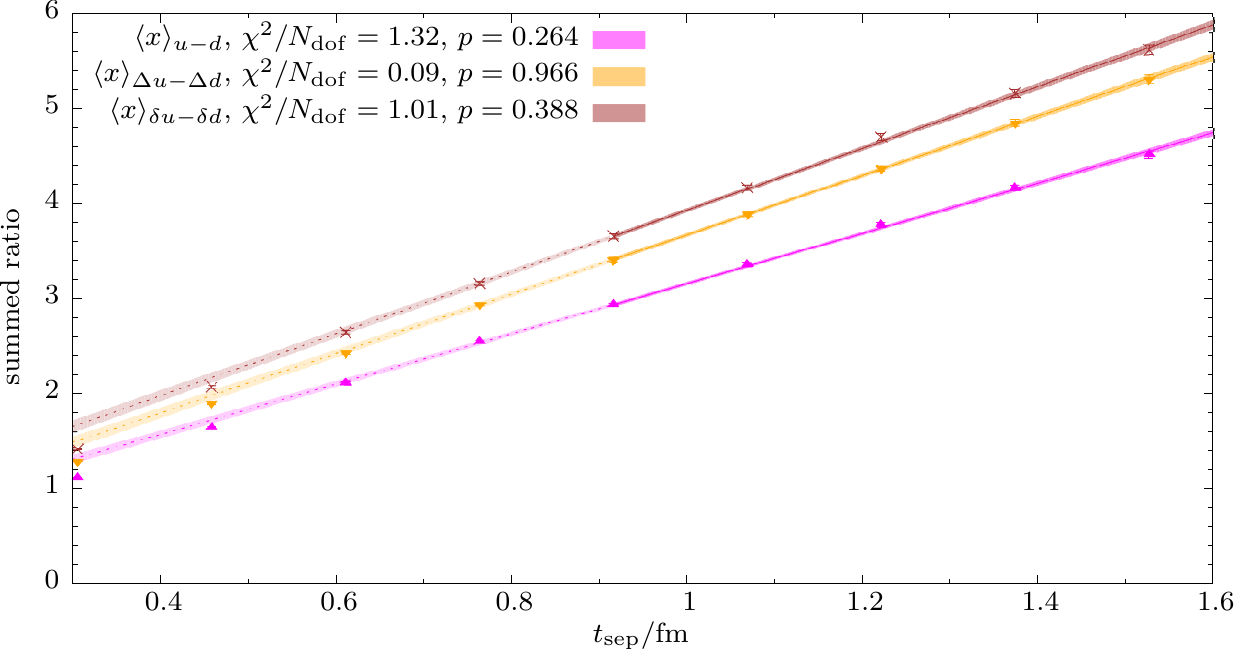}
 \includegraphics[totalheight=0.175\textheight]{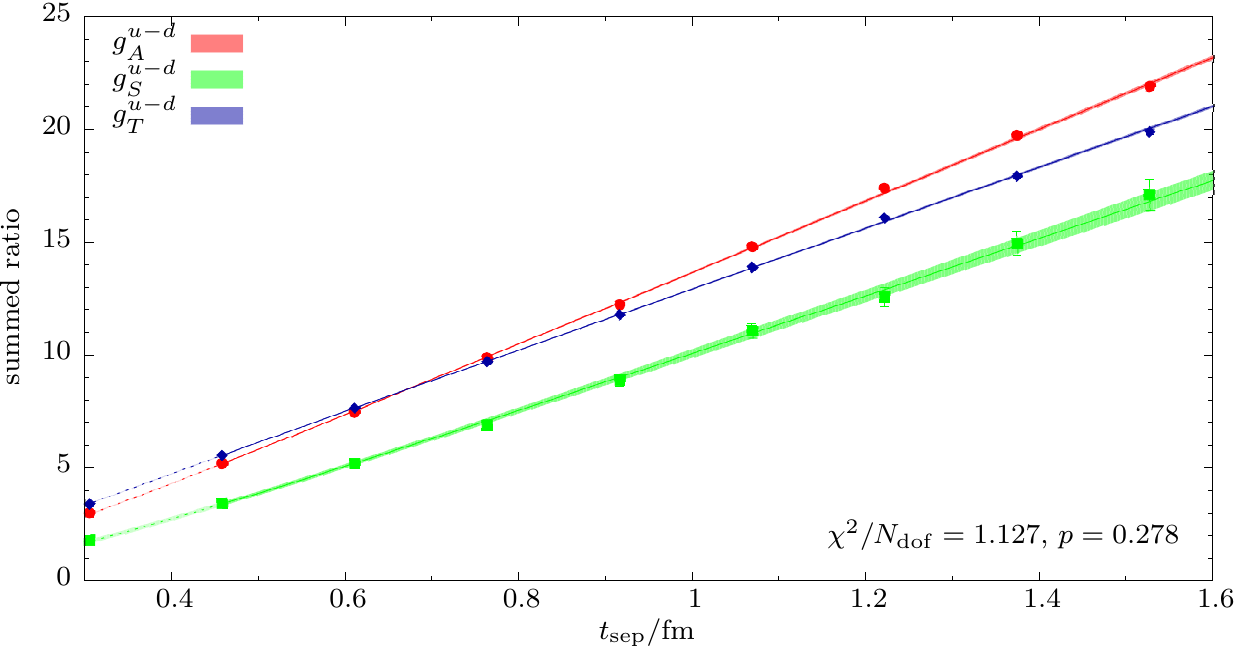}
 \includegraphics[totalheight=0.175\textheight]{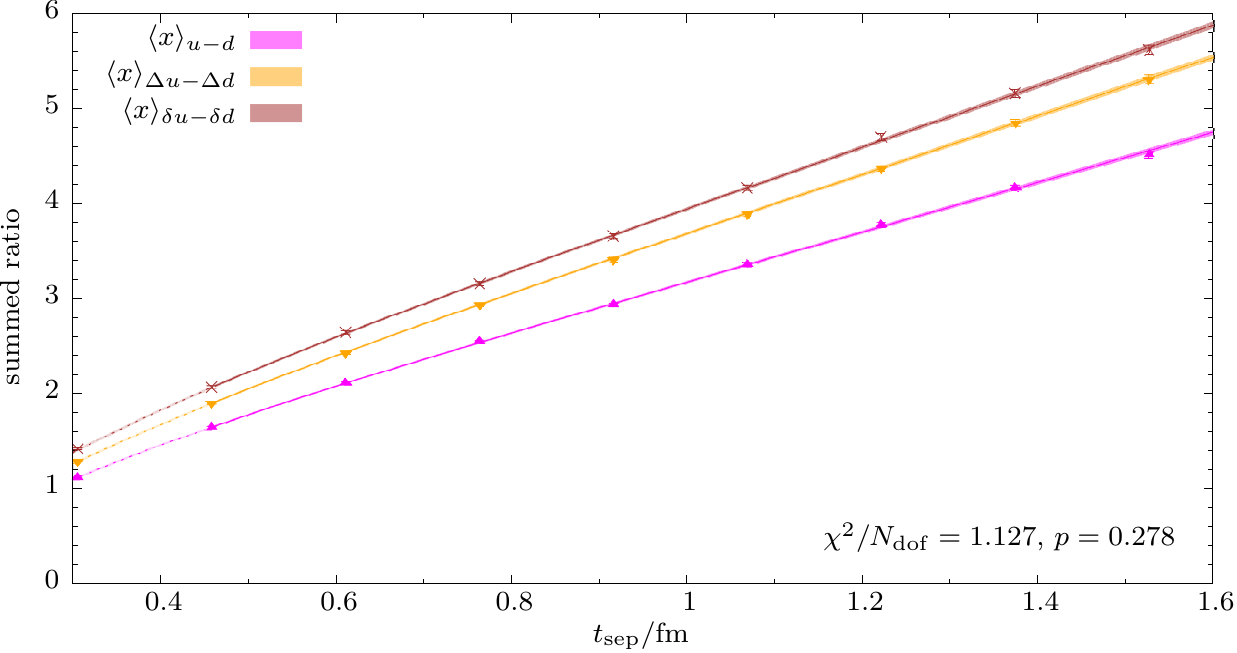}
 \caption{Exemplary fits for the single- (upper row) and two-state (lower row) summation method on the N451 ensemble, cf. Eqs.~(\ref{eq:summation_single_state}),~(\ref{eq:summation_two_state}). For the two-state summation method the corresponding fit has been performed simultaneously across all six observables. Solid lines and fit bands indicate the range of $\tsep$ values entering the fit, while dashed lines and light shaded parts of the fit bands represent an extrapolation.}
 \label{fig:summation_N451}
\end{figure}

In Fig.~\ref{fig:summation_N451} exemplary fit results are shown on N451 for all six observables, together with the resulting values of $\chi^2/\mathrm{d.o.f.}$. As expected, the lattice data deviates from the standard summation method result at small values of $\tsep$ in the upper two panels, whereas the two-state fit model in Eq.~(\ref{eq:summation_two_state}) reproduces the curvature very well across all observables. However, we observe that there is usually a drop in the resulting $p$-values below some value of $\tsepmin$, e.g. including the data at $\tsep=4a$ in Fig.~\ref{fig:summation_N451} yields $p=0.0007$ instead of $p=0.278$. We stress again, that the two-state results in the lower row are from a single, simultaneous fit which greatly stabilizes the results for individual observables. Furthermore, we find that the alternative fit model in Eq.~(\ref{eq:summation_two_state_nonuniversal}) is less stable and hence requires even smaller $\tsepmin$ values (typically $\tsepmin=4a$ or $6a$), but gives similar results otherwise. \par

\section{Physical extrapolations and outlook}
\label{sec:physical_extrapolations}
\begin{figure}[thb]
 \centering
 \includegraphics[totalheight=0.195\textheight]{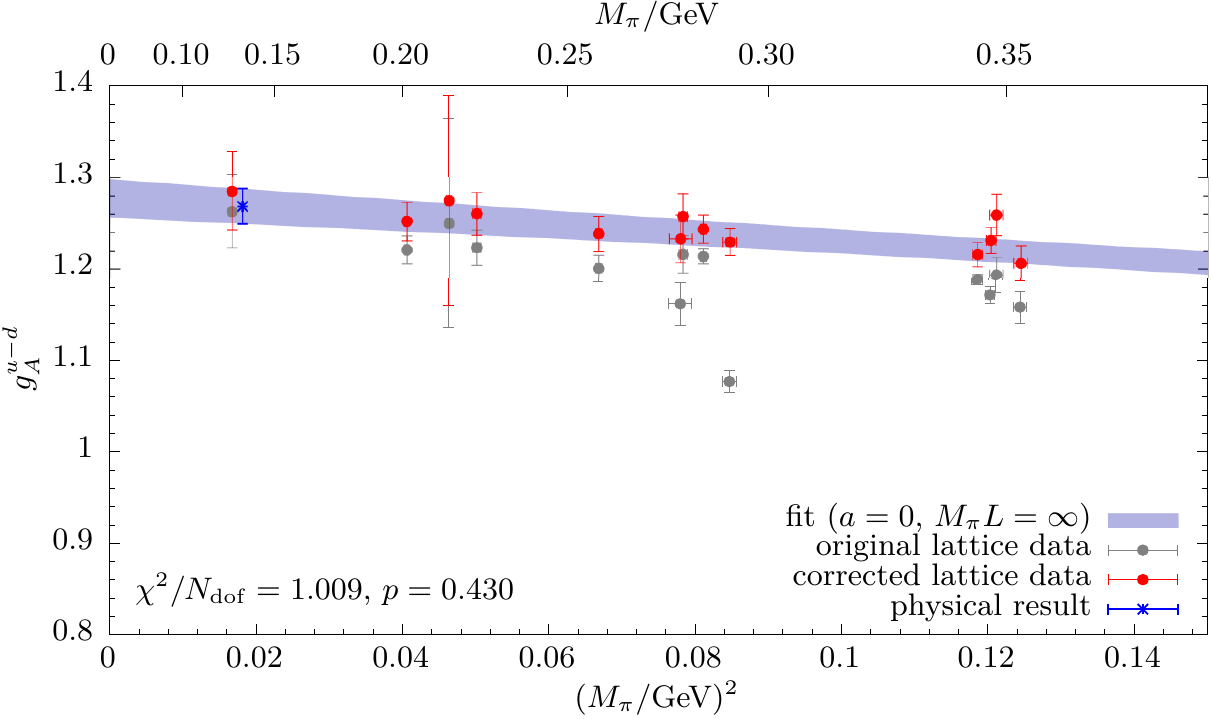}
 \includegraphics[totalheight=0.176\textheight]{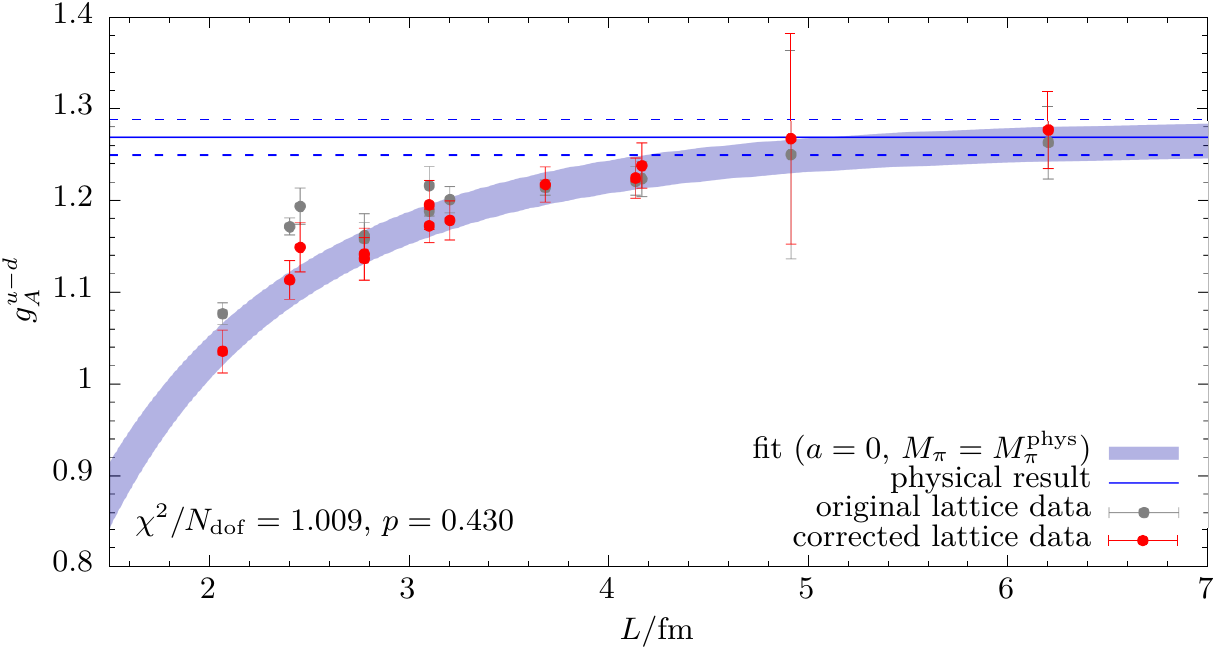} \\[3pt]
 \includegraphics[totalheight=0.195\textheight]{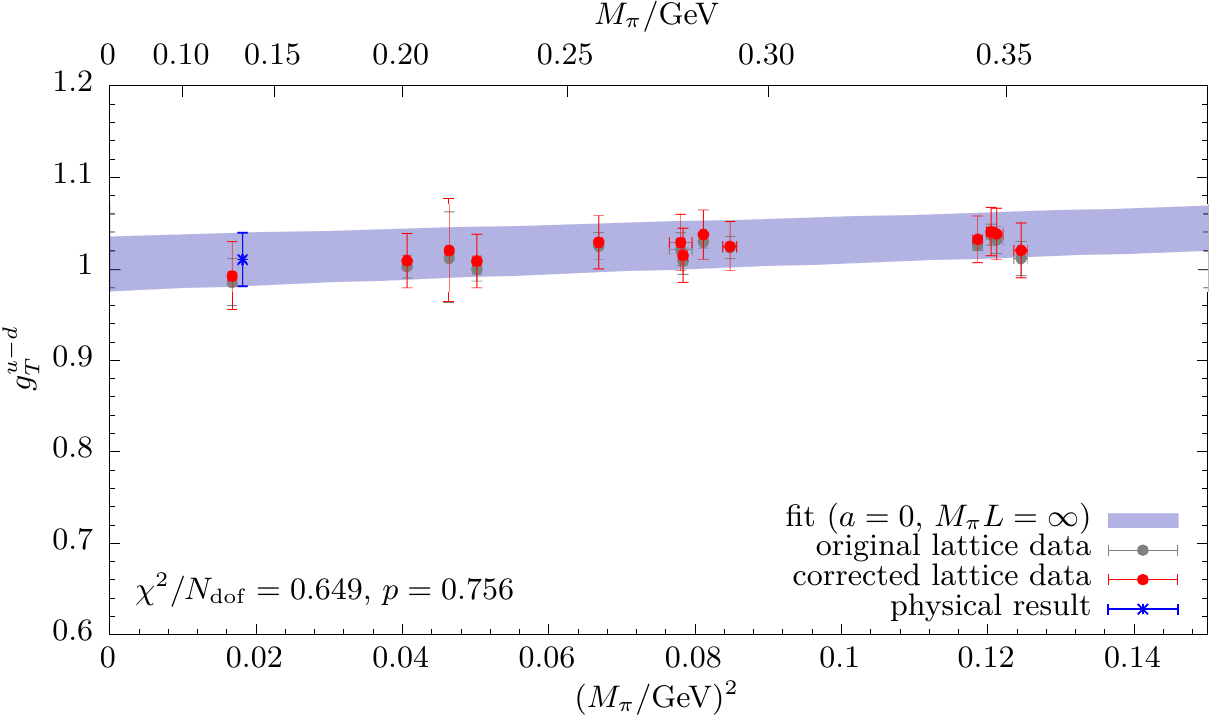}
 \includegraphics[totalheight=0.195\textheight]{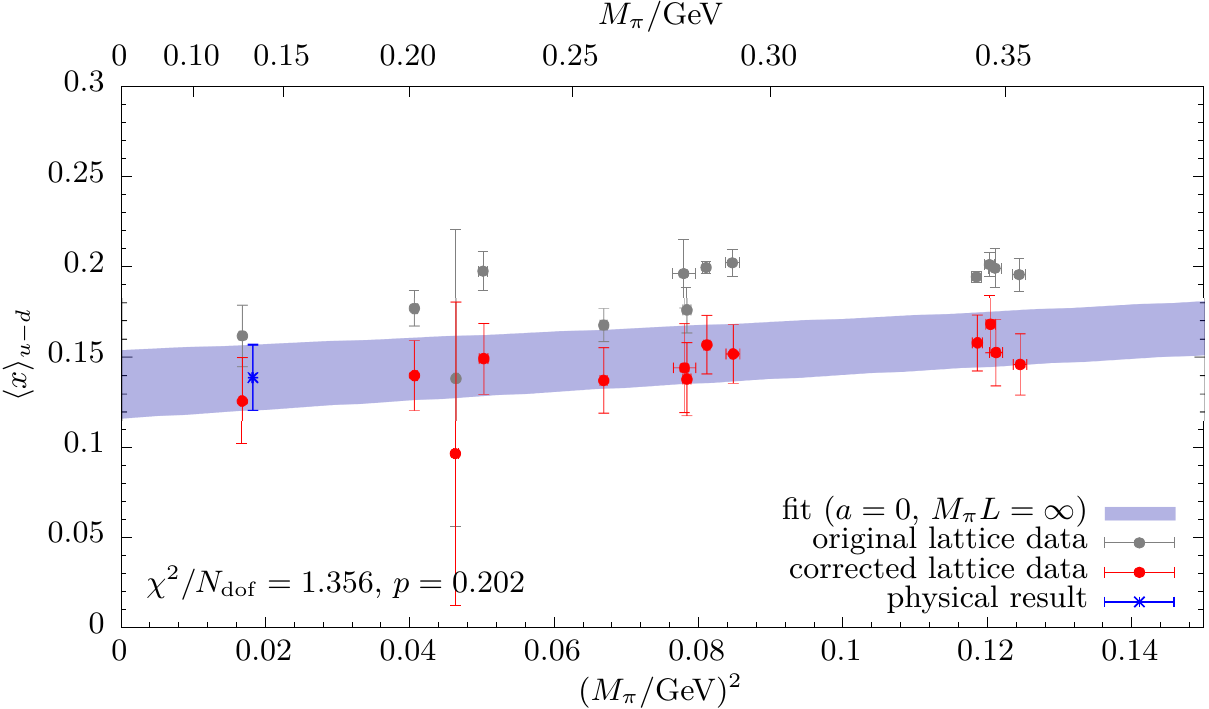}
 \caption{Chiral and finite volume extrapolation for $g_A^{u-d}$ (upper row) and chiral extrapolations for $g_T^{u-d}$ and $\avgx{-}{}$ (lower row) obtained from fitting the model in Eq.~(\ref{eq:CCF}), as described in the text. The gray symbols represent the original lattice data, whereas the red data points have been corrected for the continuum and infinite volume limit in case of the three chiral extrapolations, and for the chiral extrapolation and continuum extrapolation in case of the upper right panel. Therefore, errors on the red points are strongly correlated.}
 \label{fig:chiral_extrapolation}
\end{figure}

The physical extrapolation of the results on individual ensembles is currently performed in the same way as in Ref.~\cite{Harris:2019bih}, i.e. using an ansatz for each observable $O$ derived from a fit model inspired by chiral perturbation theory,
\begin{equation}
 O(M_\pi,a,L) = A_O + B_O M_\pi^2 + C_O M_\pi^2 \log M_\pi + D_O a^{n(O)} + E_O M_\pi^2 e^{-M_\pi L} \,. \notag
 \label{eq:CCF}
\end{equation}
The power $n(O)$ of the leading term in the continuum extrapolation is given by $n=2$ for $O=g_{A,S}^{u-d}$ and $n=1$ for all other observables. All fits are done using binned jackknife and after applying appropriate powers of the gradient flow scale $t_0/a^2$ introduced in Ref.~\cite{Luscher:2010iy} to build dimensionless quantities. For the values of $t_0/a^2$ and further information on the scale setting procedure we refer to Ref.~\cite{Bruno:2016plf} where also the physical value $\sqrt{8 t_{0,\phys}} = 0.415(4)_\stat(2)_\sys \fm$ has been determined. The latter enters our calculation only through the definition of the physical point, i.e. by fixing the physical value of $M_\pi$. The coefficient $C_O$ is known analytically for the axial charge, but fitting it as a free parameter gives the wrong sign and introduces large cancellations with the term $\sim M_\pi^2$. This is why for the time being we decided to omit the term in line with our old analysis, permitting a direct comparison of the results. For the other observables with larger statistical uncertainties the term is neglected as well. In general, the fits describe the data very well, which is reflected by the $p$-values in Fig.~\ref{fig:chiral_extrapolation}. The upper two panels show the chiral and finite-volume extrapolation for $g_A^{u-d}$ from fitting Eq.~(\ref{eq:CCF}) to the lattice data (gray symbols). Similar to what has been observed in Ref.~\cite{Harris:2019bih} the axial charge receives significant finite-volume corrections while the chiral and continuum extrapolations turn out rather mild. In the lower left panel of Fig.~\ref{fig:chiral_extrapolation} the chiral extrapolation is shown for $g_T^{u-d}$ which exhibits a flat behavior in all three extrapolations. Finally, as an example of a twist-2 matrix element the chiral extrapolation for $\avgx{-}{}$ is presented in the lower right panel. Here all three extrapolations conspire to cause a sizable downwards shift of the physical result. \par

Preliminary physical results are given in Table~\ref{tab:results} together with old results from Ref.~\cite{Harris:2019bih}. We find that the central value for $g_A^{u-d}$ is larger compared to the result from the old analysis and agrees very well with the experimental value $g_{A,\mathrm{exp}}^{u-d}=1.2724(23)$ \cite{ParticleDataGroup:2018ovx}. On the other hand, the values of the twist-2 matrix elements and in particular $\avgx{-}{}$ are further decreased. These observations are in line with the trend of the excited-state contamination observed in the effective form factors and the results are in broad agreement with experiment. In general, we find that the two-state truncation fit model in Eq.~(\ref{eq:summation_two_state}) gives results compatible with the standard summation method for the chosen constraints on $\tsepmin$, i.e. $\tsepmin M_\pi \geq0.5$ vs. $\tsepmin M_\pi \geq0.7$, respectively. It is only for $g_T^{u-d}$ that there is some tension between the two models, i.e. the value from the two-state truncation is not within the error of the standard summation method. All errors quoted in Table~\ref{tab:results} are statistical only, and they turn out similar for the two models, whereas they are reduced compared to the old analysis. \par

\begin{table}
 \centering
 \begin{tabular}{l|cccccc}
  \hline\hline
  & $g_A^{u-d}$ & $g_S^{u-d}$ & $g_T^{u-d}$ & $\langle x\rangle_{u-d}$ & $\langle x\rangle_{\Delta u-\Delta d}$ & $\langle x\rangle_{\delta u-\delta d}$ \\
  \hline
  sum. 2-state        & 1.269(19) & 1.124(96)  & 1.011(29) & 0.139(18) & 0.204(20) & 0.184(23) \\
  sum. 1-state        & 1.259(17) & 1.055(95)  & 0.969(27) & 0.136(11) & 0.193(13) & 0.182(15) \\
  \hline
  old analysis (2019) & 1.242(25) & 1.13(11)   & 0.965(38) & 0.180(25) & 0.221(25) & 0.212(32) \\
  \hline\hline
 \end{tabular}
 \caption{Preliminary, physical results for all six observables using data from a simultaneous fit of the two-state truncation fit model in Eq.~(\ref{eq:summation_two_state}), the standard summation method, and the results from the old analysis in Ref.~\cite{Harris:2019bih}. All errors are statistical only.}
 \label{tab:results}
\end{table}

In the future, we plan to assess the systematic effects related to the physical extrapolations and to revisit the aforementioned issue with the chiral logarithm at least for $g_A^{u-d}$. To this end, we intend to add further ensembles at the edges of our simulation parameter space to gain even better control for the individual terms in these fits. Moreover, we are still increasing statistics on E250, which we expect to improve the chiral extrapolation by giving more statistical weight to the most chiral data point. Concerning excited states we may further refine our analysis based on the summation method by e.g. adding the missing, intermediate (``odd'') values of $\tsep/a$ on ensembles at the coarsest lattice spacing to give a more fine grained control over $\tsepmin$. \par

\section*{Acknowledgments}
Two of the authors (GvH and KO) are supported by the Deutsche Forschungsgemeinschaft (DFG, German Research Foundation) through project HI~2048/1-2 (project No.~399400745). Calculations have been performed on the HPC clusters Clover at the Helmholtz-Institut Mainz and Mogon II and HIMster-2 at Johannes-Gutenberg Universit\"at Mainz. We gratefully acknowledge the support of the John von Neumann Institute for Computing and Gauss Centre for Supercomputing e.V. (http:www.gauss-centre.eu) for projects CHMZ21 and CHMZ36.

\end{document}

%% file: newcommands.tex
\renewcommand{\l}{\left}
\renewcommand{\r}{\right}
\newcommand{\g}[1]{\gamma_{#1}} 
\newcommand{\DB}[1]{\stackrel{\leftarrow}{D}_{#1}} 
\newcommand{\DF}[1]{\stackrel{\rightarrow}{D}_{#1}} 
\newcommand{\DBF}[1]{\stackrel{\leftrightarrow}{D}_{#1}} 

\newcommand{\bra}[1]{\left< #1 \right|} 
\newcommand{\ket}[1]{\left| #1 \right>} 

\newcommand{\gev}{\,\mathrm{GeV}}

\newcommand{\fm}{\,\mathrm{fm}}

\newcommand{\phys}{\mathrm{phys}}
\newcommand{\stat}{\mathrm{stat}}
\newcommand{\sys}{\mathrm{sys}}
\newcommand{\tex}{t_\mathrm{ex}}
\newcommand{\tins}{t_\mathrm{ins}}
\newcommand{\tinsmin}{t_\mathrm{ins}^\mathrm{min}}
\newcommand{\tsep}{t_\mathrm{sep}}
\newcommand{\tsepmin}{t_\mathrm{sep}^\mathrm{min}}

\newcommand{\tseplo}{t_\mathrm{sep}^\mathrm{lo}}
\newcommand{\tsephi}{t_\mathrm{sep}^\mathrm{hi}}

\newcommand{\avgx}[2]{\langle x \rangle_{#2 u #1 #2 d}}